\documentclass[aps,superscriptaddress,twocolumn,twoside,floatfix,pra,nofootinbib,a4paper]{revtex4}
\usepackage{times}
\usepackage{epsfig}
\usepackage{amsfonts}
\usepackage{amsmath}
\usepackage{amssymb}
\usepackage{amsthm}
\usepackage{color}
\usepackage{multirow}
\usepackage[normalem]{ulem}
\newcommand{\stkout}[1]{\ifmmode\text{\sout{\ensuremath{#1}}}\else\sout{#1}\fi}
\usepackage{latexsym}
\usepackage{mathrsfs}
\usepackage{natbib}
\usepackage{verbatim}
\usepackage[T1]{fontenc}
\usepackage{float}

\usepackage{graphicx}
\usepackage{xcolor}

\usepackage[colorlinks=true,linkcolor=blue,citecolor=magenta,urlcolor=blue]{hyperref}

\DeclareMathOperator{\Tr}{tr}
\newtheorem{lemma}{Lemma}
\newtheorem{result}{Result}

\newcommand{\ket}[1]{|#1\rangle}

\newcommand{\bracket}[3]{\langle#1|#2|#3\rangle}
\newcommand{\ketbra}[2]{|#1\rangle\langle#2|}

\begin{document}
	
	
\title{Sequential random access codes and self-testing of quantum measurement instruments}

	
\author{Karthik Mohan}
\affiliation{D\'epartement de Physique Appliqu\'ee, Universit\'e de Gen\`eve, CH-1211 Gen\`eve, Switzerland}

\author{Armin Tavakoli}
\affiliation{D\'epartement de Physique Appliqu\'ee, Universit\'e de Gen\`eve, CH-1211 Gen\`eve, Switzerland}

\author{Nicolas Brunner}
\affiliation{D\'epartement de Physique Appliqu\'ee, Universit\'e de Gen\`eve, CH-1211 Gen\`eve, Switzerland}

\begin{abstract}
Quantum Random Access Codes (QRACs) are key tools for a variety of protocols in quantum information theory. These are commonly studied in prepare-and-measure scenarios in which a sender prepares states and a receiver measures them. Here, we consider a three-party prepare-transform-measure scenario in which the simplest QRAC is implemented twice in sequence based on the same physical system. We derive optimal trade-off relations between the two QRACs. We apply our results to construct semi-device independent self-tests of quantum instruments, i.e.~measurement channels with both a classical and quantum output. Finally, we show how sequential QRACs enable inference of upper and lower bounds on the sharpness parameter of a quantum instrument.
\end{abstract}
	
		
\maketitle
	

\section{Introduction}
Random Access Codes (RACs) are an important class of communication tasks with a broad scope of applications. In a RAC, a party Alice holds a set of randomly sampled data and another party Bob attempts to recover some randomly chosen subset of Alice's data. This is made possible by Alice communicating with Bob. Therefore, this corresponds to a \textit{prepare-and-measure scenario} in which Alice encodes her data into a message that she sends to Bob who aims to decode the relevant information. Naturally, this task would be trivial if Alice is allowed to send unlimited information. Therefore, a RAC requires that the message is restricted in its alphabet, so that it cannot encode all of Alice's data. Interestingly however, the probability of Bob to access the desired information can be increased if Alice substitutes her classical message with a quantum message of the same alphabet. Such Quantum Random Access Codes (QRACs) have been introduced and developed for qubit systems \cite{Ambainis, Ozols} as well as higher-dimensional quantum systems \cite{RAC}. They are primitives for network coding \cite{NetworkCode}, random number generation \cite{HongWei} and quantum key distribution \cite{Pawlowski}. QRACs are also common in foundational aspects of quantum theory; examples include the comparison of different quantum resources \cite{TM16, HS17}, dimension witnessing \cite{Wehner}, self-testing \cite{TK18, Mate, Tavakoli} and attempts at characterising quantum correlations from information-theoretic principles \cite{IC}.

Here, we present RACs beyond standard prepare-and-measure scenarios. Specifically, we consider a `prepare-transform-measure' scenario involving three parties, Alice, Bob and Charlie, in a line configuration. In our scenario, both Bob and Charlie are interested in randomly accessing some information held by Alice, i.e.~they individually implement a RAC with Alice. In a classical picture, such sequential RACs are trivial since any information made available to Bob via Alice's communication also can be relayed by Bob to Charlie. In this sense, there is no trade-off between how well Bob and Charlie can perform their RACs. In a quantum picture however,  Alice communicates a qubit system that is first sent to Bob who applies a quantum instrument (a completely positive trace-preserving map with both a classical and quantum output) whose classical output is recorded and whose quantum output is relayed to Charlie who performs a measurement. Importantly, Bob's instrument disturbs the physical state of Alice's qubit, and therefore he  cannot relay Alice's original quantum message to Charlie. In other words, Charlie's ability to access the desired information depends on Bob's preceding interaction. Consequently, one expects a trade-off in the ability of Bob and Charlie to perform their separate QRACs.  Here, we consider Bob and Charlie the simplest RAC for qubits (sometimes referred to as a $2\rightarrow 1$ RAC) in sequence, and derive the optimal trade-off relation between the two QRACs. In particular, we find that both QRACs can outperform the best possible classical RAC.

Subsequently, we apply our results to self-test a quantum instrument. Self-testing \cite{Mayers} is the task of inferring physical entities (states, channels, measurements) solely from correlations produced in experiments i.e.~identifying the unique physical entities that are compatible with observed data. Self-testing is typically  studied in Bell experiments where notably methods for self-testing quantum instruments have been developed \cite{Wagner, Sekatski}. Recently however, self-testing was introduced in the broad scope of prepare-and-measure scenarios \cite{TK18}, and was further developed using QRACs to robustly self-test both preparations and measurements \cite{TK18, Tavakoli, Mate}. Notably however, prepare-and-measure scenarios do not enable self-tests of general quantum operations. In particular, it does not enable self-tests of quantum instruments since the quantum system after the measurement is irrelevant to the outcome statistics produced in the experiment. We show that our prepare-transform-measure scenario overcomes this conceptual limitation. We find that optimal pairs of sequential QRACs self-test quantum instruments. However, such optimal correlations require idealised (noiseless) scenarios which are never the case in a practical implementation. Therefore, we also show how sequential QRACs allow for inference of noise-robust bounds on the sharpness parameter in a quantum instrument. This is makes our results applicable to experimental demonstrations. Finally, we discuss relevant generalisations of our results.

\section{Sequential random access codes} 
\begin{figure}
	\centering
	 \includegraphics[width=\columnwidth]{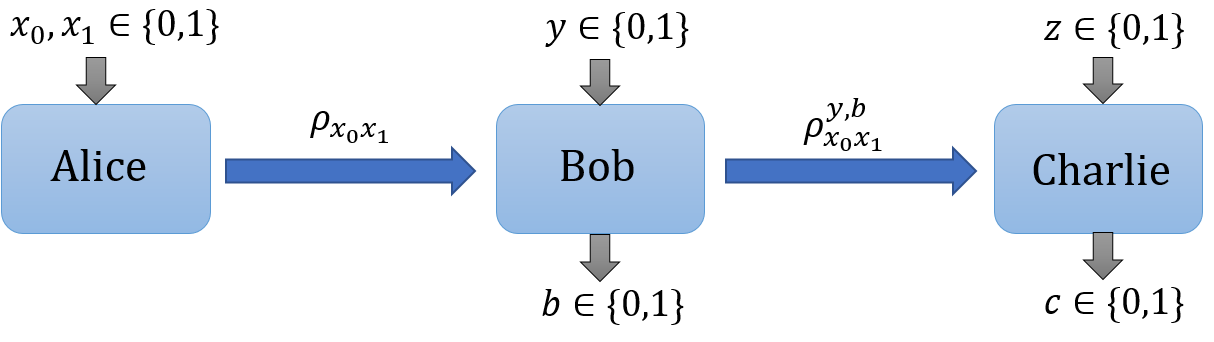}
	 \caption{A three-party prepare-transform-measure scenario. Alice samples qubit states from an ensemble of four preparations. Bob performs one of two instruments with a binary classical register and qubit output. Charlie performs one of two binary-outcome measurements.}\label{FigScenario}
\end{figure}

We focus on a prepare-transform-measure scenario that involes three parties. The first party (Alice) receives a uniformly random four-valued input $x=(x_0,x_1)\in\{0,1\}^2$. For a given input, she prepares a quantum state $\rho_x$. This state is uncharacterised, up to the assumption of it being of Hilbert space dimension two, i.e.~it is a qubit. The state is transmitted to the second party (Bob) who receives a random binary input $y\in\{0,1\}$. Depending on his input, Bob applies an instrument characterised by Kraus operators $\{K_{b|y}\}$ to $\rho_{x}$ which produces a classical binary outcome $b\in\{0,1\}$ and a qubit post-measurement state 
\begin{equation}
\rho_{x}^{y,b}=\frac{K_{b|y}\rho_xK_{b|y}^\dagger}{\Tr\left(\rho_x K_{b|y}^\dagger K_{b|y}\right)}.
\end{equation}
Notably, since the instrument realises a measurement, the Kraus operators of Bob must satisfy the completeness relation $\forall y:M_{0|y}+M_{1|y}=\openone$, where $M_{b|y}=K_{b|y}^\dagger K_{b|y}$ are the corresponding elements of the positive operator-valued measures (POVMs). The post-measurement state $\rho_x^{y,b}$ is relayed to the third party (Charlie) who receives a random binary input $z\in\{0,1\}$ to which he associates POVMs  $\{C_{c|z}\}$ with a binary outcome $c\in\{0,1\}$.  The scenario is illustrated in Figure~\ref{FigScenario}. 

In the limit of repeating the experiment many times, the results are described by the probability distribution
\begin{equation}
p(b,c|x,y,z)=\Tr\left[K_{b|y}\rho_xK_{b|y}^\dagger C_{c|z}\right].
\end{equation}
To enable a simple and qualitative treatment of the information stored in the distribution, one may employ a correlation witness, i.e.~a map from $p(b,c|x,y,z)$ to a single real number. We are interested in two separate correlation witnesses, each corresponding to a RAC. The first RAC is considered between Alice and Bob. In this task, the partners are collectively awared a point if and only if Bob can guess the $y$'th bit of Alice input $(x_0,x_1)$. The correlation witness is the average  success probability. It reads
\begin{equation}\label{WAB}
W_{\text{AB}}=\frac{1}{8}\sum_{x,y}p(b=x_y|x,y)=\frac{1}{8}\sum_{x,y} \Tr\left[\rho_x M_{x_y|y}\right],
\end{equation}
where in the second step we have assumed a quantum description. In a classical picture (in which all states are diagonal in the same basis), this witness obeys $W_\text{AB}\leq 3/4$ (which we further discuss later). The physical properties of $\{\rho_x\}$ and $\{M_{b|y}\}$ when the QRAC exceeds its classical bound were studied in Ref~\cite{TK18}. It was shown that an optimal QRAC for qubits
\begin{equation}\label{qopt}
W_\text{AB}=\frac{1}{2}\left(1+\frac{1}{\sqrt{2}}\right)\approx 0.854
\end{equation}
self-tests that Alice's four preparations form a square in some disk of the Bloch sphere. Up to a choice of reference frame these are written
\begin{align}\label{optprep}\nonumber
& \rho_{00}=\frac{1}{2}\left(\openone+\frac{\sigma_x+\sigma_z}{\sqrt{2}}\right), & \rho_{11}=\frac{1}{2}\left(\openone-\frac{\sigma_x+\sigma_z}{\sqrt{2}}\right), \\
& \rho_{01}=\frac{1}{2}\left(\openone+\frac{\sigma_x-\sigma_z}{\sqrt{2}}\right), & \rho_{10}=\frac{1}{2}\left(\openone-\frac{\sigma_x-\sigma_z}{\sqrt{2}}\right).
\end{align}
where $\vec{\sigma}=(\sigma_x,\sigma_y,\sigma_z)$ denotes the Pauli matrices. Moreover, an optimal QRAC also self-tests Bob's observables (defined as $M_y=M_{0|y}-M_{1|y}$) to be anticommuting.  In the stated frame, the observables are written 
\begin{align}\label{optmeas}
& M_0=\sigma_x & M_1=\sigma_z.
\end{align}

Evidently however, the QRAC \eqref{WAB} is independent of both Charlie and of the choice of instrument for realising the POVMs $\{M_{b|y}\}$. To also take these into account, we consider an additional QRAC implemented between Alice and Charlie. Analogously, the partners are awarded a point if and only if Charlie can guess the $z$'th bit of Alice's input $(x_0,x_1)$. The correlation witness corresponding to this QRAC reads
\begin{equation}
W_{\text{AC}}=\frac{1}{8}\sum_{x,z}p(c=x_z|x,z).
\end{equation}
This QRAC is not independent of Bob since he applies an instrument to the preparation of Alice before they arrive to Charlie. In a quantum model, the effective state $\tilde{\rho}_x$ received by Charlie is the post-measurement state of Bob averaged over Bob's inputs and classical outputs, i.e.,
\begin{equation}\label{effective}
\tilde{\rho}_x=\frac{1}{2}\sum_{y,b}p(b|y)\rho_x^{y,b}=\frac{1}{2}\sum_{y,b}K_{b|y}\rho_xK_{b|y}^\dagger.
\end{equation}
Therefore, we have
\begin{equation}
W_{\text{AC}}=\frac{1}{8}\sum_{x,z}\Tr\left[\tilde{\rho}_x C_{x_z|z}\right]=\frac{1}{16}\sum_{x,y,b,z}\Tr\left[K_{b|y}\rho_xK_{b|y}^\dagger C_{x_z|z}\right].
\end{equation}
We are interested in the values attainable for the pair of QRACs $(W_\text{AB},W_\text{AC})$. We remark that the interesting range is when both $W_\text{AB}$ and $W_\text{AC}$ are confined to the interval $[1/2, (1+1/\sqrt{2})/2]$ since either witness being $1/2-\epsilon$ for some $\epsilon>0$ is equivalent to a witness value of $1/2+\epsilon$ by classically bit-flipping the outcomes. 

Typically, we expect there to be a trade-off between the two QRACs. The reason is as follows. In order for $W_\text{AB}$ to be large, Alice must prepare states that are close to the ones in Eq~\eqref{optprep} and Bob must implement instruments that realise POVMs that are close to the ones in Eq~\eqref{optmeas}. This means that Bob's measurements must be reasonably sharp. This leads to a large disturbance in the state of the measured system which  causes the effective ensemble of states $\{\tilde{\rho}_x\}$ arriving to Charlie to lesser reflect the ensemble $\{\rho_x\}$ originally prepared by Alice. Therefore, the value of $W_\text{AC}$ is expected to be small. Conversely, if Bob makes a very unsharp measurement (almost noninteracting), he could almost completely avoid disturbing the state of Alice's system and thus we would find that $\{\tilde{\rho}_x\}$ closely approximates $\{\rho_x\}$ which allows Charlie to find a large  value of $W_\text{AC}$. However, the weak interaction of Bob then would imply a correspondingly small value of $W_\text{AB}$.

In view of the above, characterising the set of pairs $(W_\text{AB},W_\text{AC})$ that can be attained in quantum theory is a nontrivial matter. By finding such a characterisation and by understanding the trade-off between the two QRACs, we enable self-tests of Bob's instrument, along with self-tests of Alice's preparations and Charlie's measurements. Note that one may also consider alternative generalisations of QRACs to sequential scenarios \cite{Charles}.

\section{Quantum correlations in sequential random access codes}
\begin{figure}[t!]
	\centering
	\includegraphics[width=\columnwidth]{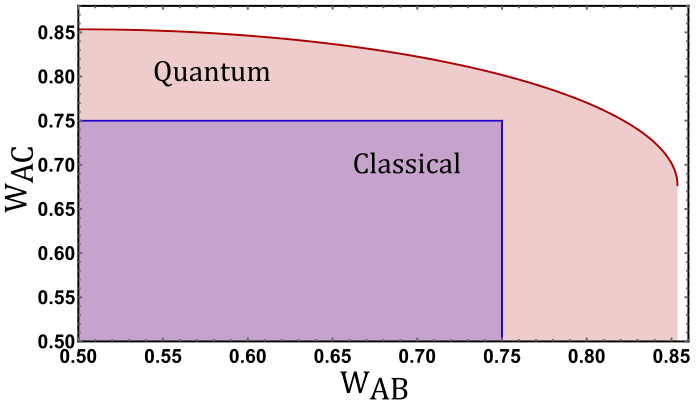}
	\caption{The correlations attainable in the space of the two witnesses $(W_\text{AB},W_\text{AC})$ in a classical and quantum model respectively. The nontrivial part of the boundary of the quantum set is highligted by a solid red line. Its right-end extremal point is $(W_\text{AB},W_\text{AC})=\left(\frac{2+\sqrt{2}}{4},\frac{4+\sqrt{2}}{8}\right)$. The extremal point for equal witnesses is $W_\text{AB}=W_\text{AC}=\frac{5+2\sqrt{2}}{10}>3/4$.}\label{FigRegion}
\end{figure}

Which values of the pair of QRACs $(W_\text{AB},W_\text{AC})$ can be realised in a quantum model based on qubit systems?  Before addressing this matter, let us first examine the substantially simpler situation in which the physical devices are classical, i.e. the state at all times is diagonal in the same basis. In such situations, Bob can interact with the preparations of Alice without disturbing their state. Therefore, a large value of $W_\text{AB}$ constitutes no obstacle for also finding a large value of $W_\text{AC}$. Classically, one can optimally achieve $W_\text{AB}=3/4$. Clearly, as the interaction with Bob cannot contribute towards increasing the value of $W_\text{AC}$, it also holds that $W_\text{AC}\leq 3/4$. This value is saturated by Alice sending $x_0$ to Bob who outputs $b=x_0$ and relays $x_0$ to Charlie who outputs  $c=x_0$. Thus, the set of classically attainable correlations is  $1/2\leq (W_\text{AB},W_\text{AC})\leq 3/4$. This classically attainable set is illustrated in Figure~\ref{FigRegion}. Notice that there is no trade-off between $W_\text{AB}$ and $W_\text{AC}$ in a classical picture.

In a quantum model, the characterisation of the attainable set of witnesses is less straightforward. We phrase the problem as follows: for a given value (denoted $\alpha$) of $W_\text{AB}$, what is the maximal value of $W_\text{AC}$ possible in a quantum model? Answering this question for every $\alpha\in[1/2,\left(1+1/\sqrt{2}\right)/2]$  provides the optimal trade-off between the two QRACs. Equivalently, it can be viewed as the nontrivial part of the boundary of the quantum set of correlations in the space of  $(W_\text{AB},W_\text{AC})$. Formally, the optimisation problem reads
\begin{align}\label{opt}\nonumber
& \qquad \qquad W_\text{AC}^\alpha=\max_{\rho,U,M,C} W_\text{{AC}} \\\nonumber
& \text{such that }  \hspace{1mm} \forall x: \rho_x\in\mathbb{C}^2, \hspace{1mm} \rho_x\geq 0,  \hspace{1mm} \Tr \rho_x=1,\\\nonumber
& \forall z,c: C_{c|z}\geq 0, \hspace{1mm} C_{0|z}+C_{1|z}=\openone\\\nonumber
& \forall y,b: U_{yb}\in\text{SU}(2), \hspace{1mm}  M_{b|y}\geq 0, \hspace{1mm} M_{0|y}+M_{1|y}=\openone,\\
& \text{and } \hspace{1mm} W_\text{AB}=\alpha,
\end{align}
i.e.~it is an optimisation of Charlie's witness over all preparations, instruments and measurements that can model the observation of $W_\text{AB}=\alpha$. In the above, we have used the polar decomposition to write the Kraus operators as $K_{b|y}=U_{yb}\sqrt{M_{b|y}}$ for some unitary operator $U_{yb}$ and some POVM $\{M_{b|y}\}$. Kraus operators of this form correspond to extremal quantum instruments in the considered scenario \cite{Pellonpää}.

We solve the problem \eqref{opt} by first giving a lower bound on $W_\text{AC}^\alpha$ and then matching it with an upper bound. To this end, consider a quantum strategy in which Alice prepares the ensemble of states given in Eq~\eqref{optprep} and Charlie performs the measurements in Eq~\eqref{optmeas}. We let Bob perform an unsharp L\"uders measurement (the Kraus operators have $U_{yb}=\openone$) of the observables in Eq~\eqref{optmeas}, i.e.~his observables correspond to 
$M_0=\eta \sigma_x$ and $M_1=\eta\sigma_z$ for some \textit{sharpness parameter} $\eta\in[0,1]$ (which we will later self-test). Evaluating the pair of witnesses with this quantum strategy gives
\begin{align}\nonumber\label{lowerbound}
& W_\text{AB}=\frac{1}{4}\left(2+\eta\sqrt{2}\right)\\
& W_\text{AC}=\frac{1}{8}\left(4+\sqrt{2}+\sqrt{2-2\eta^2}\right).
\end{align}
Parameterising the latter in terms of the former returns a lower bound on $W_\text{AC}^\alpha$. Importantly, this bound is optimal since it can be saturated with an upper bound on $W_\text{AC}^\alpha$, thus solving the optimisation problem \eqref{opt}. This leads us to our first result.

\begin{result}\label{res1}
	The optimal trade-off between the pair of QRACs $(W_\text{AB},W_\text{AC})$ corresponds to 
	\begin{equation}\label{best}
	W_\text{AC}^\alpha=\frac{1}{8}\left(4+\sqrt{2}+\sqrt{16\alpha-16\alpha^2-2}\right),
	\end{equation}
	where $\alpha\in[1/2,(1+1/\sqrt{2})/2]$. That is, the optimal witness pairs are of the form $(W_\text{AB},W_\text{AC})=(\alpha,W_\text{AC}^\alpha)$. This characterises the nontrivial boundary of the quantum set in the space of witness pairs.
\end{result}
The proof is analytical, of technical character and detailed in Appendix~\ref{AppProof}. It relies on i) treating the maximisation in \eqref{opt} over the unitaries $\{U_{yb}\}$ and measurements $\{C_{c|z}\}$ as an eigenvalue problem, ii) using the Bloch sphere parameterisation for the preparations and instruments, and iii) noticing that the maximisation over the preparations can be relaxed to a maximisation over two pairs of antipodal pure states in some disk of the Bloch sphere.

In Figure~\ref{FigRegion}, we have illustrated the set of sequential QRACs attainble in a quantum model.  Notice that a maximal value \eqref{qopt} of $W_\text{AB}$ does not imply that $W_\text{AC}$ is no better than what is obtained by random guessing. In contrast, one can achieve $(W_\text{AB},W_\text{AC})=\left(\frac{2+\sqrt{2}}{4},\frac{4+\sqrt{2}}{8}\right)$. The reason is that the ensemble relayed to Charlie corresponds to that originally prepared by Alice but with Bloch vectors of half the original length. In addition, there exists a subset of the quantum set in which both $W_{\text{AB}}$ and $W_\text{AC}$ exceed the classical bound.

\section{Self-testing}
Finding the optimal trade-off between the two QRACs (Result~\ref{res1}) allows for self-testing. To obtain a self-test, one must additionally show that the optimal QRAC pairs only admit a realisation with unique preparations, instruments and measurements (up to collective unitary transformations). That is, we need to identify the unique physical entities $\{\rho_x\}$, $\{K_{b|y}\}$, and $\{C_{c|z}\}$ necessary for optimal correlations. 

Such a self-testing argument can be established largely from the proof of Result~\ref{res1} (see Appendix~\ref{AppProof}). The reason is that our approach to deriving Result~\ref{res1} successively identifies the form of the physical entities required for optimality. To turn the statement into a self-test, we identify key inequalities used to upper bound $W_\text{AC}^\alpha$ and instead impose strict equality constraints. This allows us to pinpoint the states, measurements and instruments one by one. These additional arguments are discussed in Appendix~\ref{AppProof}.  This leads us to the following self-test statement based on optimal sequential QRACs.

\begin{result}\label{res2}
	An optimal pair of QRACs $(W_\text{AB},W_\text{AC})=(\alpha,W_\text{AC}^\alpha)$, as in Eq.~\eqref{opt}, self-tests that
	\begin{itemize}
		\item Alice's states are pure and pairwise antipodal on the Bloch sphere, on which they form a square. These correspond to the states given in Eq.~\eqref{optprep}.
		\item Bob's instruments are Kraus operators  $K_{b|y}=U_{yb}\sqrt{M_{b|y}}$ that correspond to unsharp measurements along the diagonals of Alice's square of preparations followed by a collective unitary. Specifically, $\forall y,b: U_{yb}=U$, $M_0=\eta\sigma_x$ and $ M_1=\eta\sigma_z$ where $		\eta=\sqrt{2}\left(2W_\text{AB}-1\right)$.
		\item Charlie's measurements are rank-one projective along the diagonals of the square formed by Alice's preparations, up to the unitary of Bob. That is, $C_0=U\sigma_xU^\dagger$ and $C_1=U\sigma_zU^\dagger$.
	\end{itemize} 
	The self-tests are valid up to a collective choice of reference frame.
\end{result}

This result applies to optimal pairs of QRACs (highlighted by a solid red line in Figure~\ref{FigRegion}). An interesting question is how to make this result noise-tolerant so that it applies to suboptimal pairs of QRACs that nevertheless lack a classical model. Naturally, when the QRACs are suboptimal, one can no longer pinpoint the physical entities as done in Result~\ref{res2}. However, it is possible to give qualitative statements about the quantum strategies that in principle could model the observed correlations. We consider this matter for the sharpness parameter in Bob's instruments. Since any binary-outcome qubit observable can be written on the form $M_y=c_{y0}\openone+\vec{c}_y\cdot \vec{\sigma}$, we define the sharpness parameter of Bob's instrument as the length of the Bloch vector $\vec{c}_y$. For simplicity, we take both his instruments to have the same sharpness $\eta\equiv |\vec{c}_0|=|\vec{c}_1|$.

We can place a lower bound on $\eta$ from the witness $W_\text{AB}$; it corresponds to the smallest $\eta$ for which there exists preparations and instruments that can model $W_\text{AB}$. In Appendix~\ref{AppB}, we show that this lower bound reads 
\begin{equation}\label{lb}
\eta\geq \sqrt{2}\left(2W_\text{AB}-1\right).
\end{equation}
This lower bound is nontrivial whenever $W_\text{AB}>1/2$. Notice also that an optimal QRAC \eqref{qopt} necessitates a sharp measurement ($\eta=1$). Similarly, we can place an upper bound on $\eta$ from the witness $W_\text{AC}$, corresponding to the largest $\eta$ for which there exists preparations, instruments and measurements that can model $W_\text{AC}$. In Appendix~\ref{AppB} we show that such a bound reads 
\begin{equation}\label{ub}
\eta \leq 2\sqrt{\left(2+\sqrt{2}-4W_\text{AC}\right)\left(2W_\text{AC}-1\right)},
\end{equation}
when $\frac{4+\sqrt{2}}{8}\leq W_\text{AC}\leq \frac{2+	\sqrt{2}}{4}$ (otherwise the bound is trivial). The lower bound \eqref{lb} and the upper bound \eqref{ub} are tight, i.e.~they can be saturated with an explicit quantum strategy. Notice that the upper bound \eqref{ub} conincides with the lower bound \eqref{lb} for optimal $W_\text{AC}$ (i.e.~when $W_\text{AC}=W_\text{AC}^\alpha$) as given in Eq~\eqref{best}. In addition, the bound \eqref{ub} reduces to the trivial $\eta\leq 1$ when $W_\text{AC}=\left(4+\sqrt{2}\right)/8\approx 0.6767$. 

As a simple example, consider an experiment that attempts to implement the quantum strategy \eqref{lowerbound} for the optimal witness pair $(W_\text{AB},W_\text{AC})$ corresponding to $\eta=1/\sqrt{2}$. However the experiment is subject to losses. For example, take a $95\%$ visibility\footnote{Here, visibility corresponds to a parameter $v\in[0,1]$ and means that the ideal physical entity is implemented with probability $v$ and with probability $(1-v)$ the implemented physical entity is maximally mixed.} in Alice's preparations, $90\%$ visibility in Bob's instruments, and $95\%$ visibility in Charlie's measurements. Instead of finding the optimal witness pair $(W_\text{AB},W_\text{AC})=(3/4,(5+\sqrt{2})/8)$, one finds $(W_\text{AB},W_\text{AC})\approx(0.7138,0.7826)$. Therefore, we find that $\eta$ must be confined to the interval $0.6047\leq \eta \leq 0.8010$.  The interval is fairly wide, which emphasises the need for high-quality practical realisations in order to confine $\eta$ to a reasonably small interval.

\section{Generalisations}
Above, we have thoroughly considered the scenario in which a sequence of three observers implement a pair of the simplest  QRAC. This is arguably the simplest scenario in which to study sequential QRACs. It would be interesting to consider more general scenarios; both involving higher-dimensional \cite{RAC} and many-input QRACs, as well as sequences of more than three observers. 

Consider for example the above considered RAC played between Alice and a sequence of $N$ parties. We denote the RAC between Alice and sequential party number $k$  by $W_{k}$. Let Alice prepare the optimal states in Eq.~\eqref{optprep}. We know that if the first party performs optimal projective measurements \eqref{optmeas} (with Kraus operators $K_{b|y}=M_{b|y}$), he will find the optimal QRAC given in Eq.~\eqref{qopt}. Moreover, if the second party performs the same Kraus operators we find $W_3=\left(1+1/(2\sqrt{2})\right)/2$. The reason is that the effective state ensemble \eqref{effective} relayed by the first party is identical the the preparations of Alice except that their Bloch vectors have shrunk to half the unit lenght. Similarly, the effective ensemble relayed by the second party will be identical to that relayed by the first party, except that the Bloch vectors will again by shrunk to a quarter of unit length. Continuing the sequence in this manner, the square formed in the Bloch sphere by the effective post-measurement ensemble will at each step have its half-diagonal reduced by a factor $1/2$, and we find 
\begin{equation}
W_k=\frac{1}{2}\left(1+\frac{\sqrt{2}}{2^{k}}\right).
\end{equation}

Moreover, one can ask what is the longest sequence of QRACs such that all of them can exceed the classical bound. The number is at least two, since we found $W_\text{AB}=W_\text{AC}=\frac{5+2\sqrt{2}}{10}\approx 0.7828>3/4$. However, a third sequential violation is unlikely to be possible, i.e.~to find  $W_1=W_2=W_3>3/4$. The reason is based on the possibility of relating witnesses in dimension-bounded prepare-and-measure scenarios to Bell inequalities \cite{Buhrmann, BZ04, TZ17}. Via such methods, the considered RAC can be related to the CHSH inequality \cite{Buhrmann}. However, sequential violations of the CHSH inequality were studied in Ref.~\cite{Ralph} and it was found that no more than two CHSH inequality violations are possible when inputs are uniformly distributed \cite{Shenoy, Dipankar}.

\section{Conclusions}
We have studied sequential Quantum Random Access Codes and characterised their optimal trade-off. This ties in with the recent interest in sequential quantum correlations obtained in various forms of tests of nonclassicality \cite{Curchod, Gallego, Sasmal, Bera, Shenoy, TC18, Ralph, Wilson}. We applied our results to show that quantum instruments can be semi device-independently self-tested.  Notably, since all quantum instruments also realise some POVM, our results trivially implies a certification of unsharp measurements. Our results complement the many recent self-tests of preparations and measurements in standard prepare-and-measure scenarios with a method for self-testing quantum instruments. In addition, we showed how to robustly certify the sharpness parameter of quantum instruments based on noisy correlations. This makes our results readily applicable to experimental applications. Such tests are well within the state-of-the-art experiments \cite{Exp1, Exp2, Wilson}. Moreover, we notice that the class of quantum instruments self-tested in this work are precisely those implemented by the experimental realisations in  Ref.~\cite{Exp1, Exp2, Wilson}. 

We conclude with some open questions. Firstly, it would be interesting to generalise our results to cover higher-dimensional QRACs and longer sequences of observers. Secondly, a possible further development is to characterise the optimal trade-off between sequential QRACs encountered in tests of preparation contextuality \cite{Wilson}. Thirdly, in the spirit of Ref.~\cite{Wagner}, it would be interesting to develop noise-robust self-testing of quantum instruments. Typically, such a robust self-test address the closeness (based on observed witness values) between the unknown laboratory instrument and the ideal instrument that would have been self-tested in case correlations were optimal. Finally, one could consider the task of self-testing quantum instruments based on the sequential correlation experiments in the fully device-independent scenario (see Ref.~\cite{Ralph}).

\textit{Note added.---} During the completion of this work, we became aware of the related work of Ref.~\cite{Miklin}.

\section{Acknowledgements} 
We thank Denis Rosset and J\k{e}drzej Kaniewski for discussions. This work was funded by the Swiss National Science Foundation (Starting grant DIAQ, NCCR-QSIT).

\appendix

\section{Proof of Result~\ref{res1} and Result~\ref{res2}}\label{AppProof}
We first prove Result~\ref{res1} and then develop the argument further to also prove Result~\ref{res2}.

Consider the maximisation of the witness
\begin{equation}\label{e1}
W_\text{AC}=\frac{1}{16}\sum_{x,y,b,z}\Tr\left[K_{b|y}\rho_x K_{b|y}^\dagger C_{x_z|z}\right]
\end{equation}
under the constraint that 
\begin{equation}\label{e2}
\alpha \equiv W_\text{AB}= \frac{1}{8}\sum_{x,y}\Tr\left[\rho_xK_{x_y|y}^\dagger K_{x_y|y}\right].
\end{equation}
The optimisation is relevant for every $\alpha\in\left[1/2,\left(1+1/\sqrt{2}\right)/2\right]$, ranging from the trivial witness value to the maximal witness value.

To contend with this, we first use the polar decomposition $K_{b|y}=U_{yb}\sqrt{M_{b|y}}$, where $U_{yb}$ are arbitrary unitary operators. We can then use the cyclicity of the trace along with the substitution $C_{1|z}=\openone-C_{0|z}$ to write Eq~\eqref{e1} as
\begin{equation}\label{step1}
W_\text{AC}=\frac{1}{2}+\frac{1}{16}\sum_{x,y,b,z}(-1)^{x_z}\Tr\left[\sqrt{M_{b|y}} \rho_x \sqrt{M_{b|y}} U_{yb}^\dagger C_{0|z}U_{yb}\right].
\end{equation}
The sum over $x$ can be moved inside the trace; we define $\gamma_z=\sum_x (-1)^{x_z}\rho_x$. Moreover, we also define $A_{zyb}=U_{yb}^\dagger C_{0|z}U_{yb}$. We can now consider the optimisation over $\{U_{yb}\}$ and $\{C_{c|z}\}$ as a single optimisation over $A_{zyb}$. To this end, we note that the set of measurements $\{C_{c|z}\}$ is convex. Therefore, every nonextremal (interior point) measurement can be written as a convex combination of extremal measurements (on the boundary). Due to linearity, no nonextremal POVM can lead to a larger value of $W_\text{AC}$ than some extremal POVM. The extremal binary-outcome qubit measurements are rank-one projectors. Therefore, we can consider the optimisation over $A_{zyb}$ as an optimisation over general rank-one projectors.  This gives 
\begin{multline}\label{step2}
\max W_\text{AC}=\frac{1}{2}+\max_{\rho,A,M}\frac{1}{16}\sum_{y,b,z}\Tr\left[\sqrt{M_{b|y}} \gamma_z \sqrt{M_{b|y}}A_{zyb}\right]\\
=\frac{1}{2}+\max_{\rho,M}\frac{1}{16}\sum_{y,b,z}\lambda_{\text{max}}\left[\sqrt{M_{b|y}} \gamma_z \sqrt{M_{b|y}}\right],
\end{multline}
where we have made the optimal choice of letting $A_{zyb}$ project onto the eigenvector of $\sqrt{M_{b|y}} \gamma_z \sqrt{M_{b|y}}$ with the largest eigenvalue (denoted by $\lambda_{\text{max}})$.

To proceed further, we make use of the fact that qubit operations can be parameterised on the Bloch sphere. We write the preparations as $\rho_x=\left(\openone+\vec{n}_x\cdot\vec{\sigma}\right)/2$ for some Bloch vectors $\vec{n}_x\in\mathbb{R}^3$ with $|\vec{n}_x|\leq 1$. This leads to 
\begin{equation}
\gamma_z=\left[\left(\vec{n}_{00}-\vec{n}_{11}\right)+(-1)^z\left(\vec{n}_{01}-\vec{n}_{10}\right)\right]\cdot \vec{\sigma}.
\end{equation}
We define the effective (unnormalised) Bloch vectors $\vec{m}_z=\left(\vec{n}_{00}-\vec{n}_{11}\right)+(-1)^z\left(\vec{n}_{01}-\vec{n}_{10}\right)$. Consequently, the dependence of $W_\text{AC}$ on the preparations can be reduced to its dependence on $(\vec{m}_0,\vec{m}_1)$. However, given any set of preparations $\{\vec{n}_x\}$, we can consider other preparations  $\{\vec{n}'_x\}$ choosen such that $\vec{n}'_{00}=-\vec{n}'_{11}$ and $\vec{n}'_{01}=-\vec{n}'_{10}$ with $2\vec{n}'_{00}= \vec{n}_{00}-\vec{n}_{11}$ and $2\vec{n}'_{01}=\vec{n}_{01}-\vec{n}_{10}$. The both ensembles $\{\vec{n}_x\}$ and $\{\vec{n}'_x\}$  imply the same vectors $(\vec{m}_0,\vec{m}_1)$. Moreover, it is evident that if not all preparations are pure, one cannot obtain optimal correlations (since impurity corresponds to decreasing the magnitude of $(\vec{m}_0,\vec{m}_1)$). This means that the Bloch vectors are of unit lenght and therefore that the optimal preparations \textit{must} be of the type $\{\vec{n}_x'\}$ (i.e.~two antipodal pairs). Notice that purity also implies that $\vec{m}_0\cdot \vec{m}_1=0$.  

W.l.g. we can choose a reference frame in which $\vec{m}_0\propto (1,0,0)$ and $\vec{m}_1\propto (0,0,1)$. We denote the relative angle between the two pairs of antipodal preparation pairs by $\theta\in[0,\pi/2]$. This gives 
\begin{align}\nonumber
& |\vec{m}_0|=\sqrt{2(1+\cos\theta)} & \text{and} &&  |\vec{m}_1|=\sqrt{2(1-\cos\theta)}.
\end{align} 

We can further place an upper bound on Eq~\eqref{step2} by using the following relation
\begin{equation}\label{lemma2}
\forall M, \forall \vec{a}\in\mathbb{R}^3:\hspace{1mm} \sum_{b=0,1}\lambda_{\text{max}}\left[\sqrt{M_b}(\vec{a}\cdot\vec{\sigma})\sqrt{M_b}\right]\leq |\vec{a}|,
\end{equation}
with equality if and only if $\vec{a}$ is aligned with the Bloch vector of the POVM. Identifying $\vec{a}$ with $\vec{m}_z$, we apply it  twice to Eq~\eqref{step2} corresponding to the terms in which $z=y$. This gives
\begin{multline}\label{rrr}
W_\text{AC}\leq \\
\frac{1}{2}+\frac{1}{16}\bigg(|\vec{m}_0|+|\vec{m}_1|+\underbrace{\sum_{y,b} \lambda_{\text{max}}\left[\sqrt{M_{b|y}} (\vec{m}_{\bar{y}}\cdot\vec{\sigma}) \sqrt{M_{b|y}}\right]}_{=S}\bigg),
\end{multline}
where $\bar{y}$ denotes a bit-flip. We turn our attention to the sum denoted by $S$ in Eq~\eqref{rrr}. We define the observable $M_y=M_{0|y}-M_{1|y}$ and apply the Bloch sphere parameterisation. We may write  $M_y=c_{y0}\openone+\vec{c}_y\cdot\vec{\sigma}$ where $\vec{c}_y=(c_{y1},c_{y2},c_{y3})$ with $|\vec{c}_y|\leq 1$ and $|\vec{c}_y|-1\leq c_{y0}\leq 1-|\vec{c}_y|$. These constraints ensure positivity. Hence, 
\begin{equation}
M_{b|y}=f_{yb}\ketbra{\vec{c}_y}{\vec{c}_y}+h_{yb}\ketbra{-\vec{c}_y}{-\vec{c}_y}
\end{equation}
where $\ket{\vec{c}_y}$ is the pure state corresponding to the Bloch sphere direction of $\vec{c}_y$, and 
\begin{align}\label{fh}\nonumber
& f_{yb}=\frac{1}{2}\left(1+(-1)^bc_{y0}+(-1)^b|\vec{c}_y|\right)\\
& h_{yb}=\frac{1}{2}\left(1+(-1)^bc_{y0}-(-1)^b|\vec{c}_y|\right).
\end{align}
Firstly, this allows us to write the constraint \eqref{e2} as
\begin{equation}\label{cc}
\alpha=\frac{1}{8}\left(4+|\vec{m}_0| c_{01}+|\vec{m}_1| c_{13}\right).
\end{equation}
Secondly, we can now solve the characteristic equation $\det\left(\sqrt{M_{b|y}} (\vec{m}_{\bar{y}}\cdot\vec{\sigma}) \sqrt{M_{b|y}}-\mu\openone\right)=0$, and after  some simplifications obtain 
\begin{multline}\label{tt}
S=\sum_{y,b}\frac{|\vec{m}_{\bar{y}}|}{2} \\
\times \sqrt{\left(1+(-1)^bc_{y0}\right)^2-|\vec{c}_y|^2\left(1-\bracket{\vec{c}_y}{\hat{m}_{\bar{y}}\cdot\vec{\sigma}}{\vec{c}_y}^2\right)},
\end{multline}
where $\hat{m}=\vec{m}/|\vec{m}|$. We can now consider the optimisation over $c_{y0}$ by separately considering the two terms corresponding to $y=0$ and $y=1$ respectively. This amounts to maximising expressions of the form $\sqrt{(1+x)^2-K}+\sqrt{(1-x)^2-K}$, for some positive constant $K$. It is easily shown that such functions are uniquely maximised by setting $x=0$. Thus, we require $c_{00}=c_{10}=0$. Moreover, since $(\vec{m}_0,\vec{m}_1)$ have no component along the $y$-axis, it is seen from \eqref{cc} and \eqref{tt} that one optimally chooses $c_{02}=c_{12}=0$. This simplifies matters to 
\begin{multline}\label{S}
\max S=|\vec{m}_0|\sqrt{1-(c_{11}^2+c_{13}^2)\left(1-c_{11}^2\right)}\\
+|\vec{m}_1|\sqrt{1-(c_{01}^2+c_{03}^2)\left(1-c_{03}^2\right)}
\end{multline}
Note that $c_{03}$ and  $c_{11}$ do not appear in the constraint \eqref{cc},  that they are associated to different settings of Bob and that they appear in different terms in Eq~\eqref{S}. Therefore, we can separately maximise search square-root expression above by standard differentiation. This returns that the unique maximum is attained for $c_{03}=c_{11}=0$. Hence, we have  
\begin{multline}
W_\text{AC}\leq \frac{1}{2}+\frac{1}{16}\bigg(|\vec{m}_0|+|\vec{m}_1|\\
+|\vec{m}_0|\sqrt{1-c_{13}^2}+|\vec{m}_1|\sqrt{1-c_{01}^2}\bigg)\equiv W
\end{multline}
Denoting $c_{01}=\cos\phi_0$ and $c_{13}=\cos\phi_1$ for $\phi_1,\phi_2\in[0,\pi/2]$, we can re-write the right-hand-side on the more convenient form
\begin{equation}\label{objtrig}
W=\frac{1}{2}+\frac{1}{8}\bigg(\cos\frac{\theta}{2}+\sin\frac{\theta}{2}+\cos\frac{\theta}{2}\sin\phi_1+\sin\frac{\theta}{2}\sin\phi_0
\bigg)
\end{equation}
and the constraint \eqref{cc} as
\begin{equation}\label{constrig}
\alpha=\frac{1}{8}\left(4+\cos\frac{\theta}{2}\cos\phi_0+\sin\frac{\theta}{2}\cos\phi_1\right).
\end{equation}
To maximise $W$ over $(\theta,\phi_0,\phi_1)$, we use the following lemma.

\begin{lemma}\label{lemma}
	For every tuple $(\theta,\phi_0,\phi_1)$ corresponding to $(\alpha, W)$, there exists another tuple $(\theta,\phi_0,\phi_1)=(\pi/2,\phi,\phi)$ that produces $(\alpha,W')$ with $W'\geq W$. Moreover, $\theta=\pi/2$ and $\phi_0=\phi_1$ is necessary for an optimal $W'$. 
\end{lemma}
To prove this statement, we must show that for all $\theta,\phi_0,\phi_1\in[0,\pi/2]$ there exists a $\phi\in[0,\pi/2]$ such that
\begin{align}\label{lemmaeq}\nonumber
& \cos\frac{\theta}{2}\cos\phi_0+\sin\frac{\theta}{2}\cos\phi_1=\sqrt{2}\cos\phi\\
& \cos\frac{\theta}{2}+\sin\frac{\theta}{2}+\cos\frac{\theta}{2}\sin\phi_1+\sin\frac{\theta}{2}\sin\phi_0\leq\sqrt{2}+\sqrt{2}\sin\phi.
\end{align}
\begin{proof}
It trivially holds that $\cos\frac{\theta}{2}+\sin\frac{\theta}{2}\leq \sqrt{2}$ with equality if and only if $\theta=\pi/2$. We eliminate this part from the second equation in \eqref{lemmaeq}. Then, by squaring both equations, we can combine them into a single equation in which $\phi$ is eliminated. The statement reduces to the inequality
\begin{equation}\label{fin}
\cos\theta\left(\cos^2\phi_0-\cos^2\phi_1\right)+\sin\theta\cos\left(\phi_0-\phi_1\right)\leq 1.
\end{equation}
Using differentiation w.r.t. $\phi_0$ one finds that the optimum of the left-hand-side is attained for $\phi_1=\phi_0$, which proves the relation \eqref{fin}.
\end{proof}

By virtue of lemma~\ref{lemma}, we can reduce our consideration of \eqref{objtrig} and \eqref{constrig} to $\theta=\pi/2$ and $c_{01}=c_{13}\equiv c$. Therefore Eq~\eqref{cc} reduces to 
\begin{equation}\label{length}
c=\sqrt{2}\left(2\alpha-1\right)
\end{equation}
and we also have $W=\frac{1}{2}+\frac{1}{4\sqrt{2}}\left(1+\sqrt{1-c^2}\right)$. Thus, we have arrived to the upper bound
\begin{equation}\label{fineq}
 W_\text{AC}^\alpha\leq  \frac{1}{8}\left(4+\sqrt{2}+\sqrt{16\alpha-16\alpha^2-2}\right).
\end{equation}
As shown in the main text, this upper bound could be saturated with an explicit quantum strategy. This proves Result~\ref{res1}.

Let us now extend this to a proof of Result~\ref{res2} by more closely examining the above steps needed to arrive at Eq~\eqref{fineq}. Firstly, we have already shown that the preparations must be pure, pairwise antipodal and by lemma~\ref{lemma} they must have a relative angle of $\pi/2$. Thus, this corresponds to a square in a disk of the Bloch sphere. The above arguments fully characterise Alice's preparations up to a reference frame. 

For Bob's instrument, we have shown that the Bloch vectors $(\vec{c}_0,\vec{c}_1)$ only can have non-zero components in the $x$- and $z$-directions respectively and that the length of the Bloch vector is given by Eq~\eqref{length}. This fully characterises the Bloch vectors. Moreover, in Eq~\eqref{step2} we required that $A_{zyb}$ is aligned with the eigenvector of $\sqrt{M_{b|y}}\gamma_z\sqrt{M_{b|y}}$ corresponding to the largest eigenvalue. However, we now have that $\gamma_0=\sigma_x$ and $\gamma_1=\sigma_z$ whereas $M_0\propto \sigma_x$ and $M_1\propto \sigma_z$. Therefore, we have that $\forall y,b: A_{0yb}=\ketbra{+}{+}$ and $\forall y,b: A_{1yb}=\ketbra{0}{0}$. Therefore, we have that 
\begin{align}
&\forall yb: & U^\dagger_{yb} C_{0|0} U_{yb}=\ketbra{+}{+} \\
&\forall yb: & U^\dagger_{yb} C_{0|1} U_{yb}=\ketbra{0}{0} .
\end{align}
This implies that all unitaries are equal; $U_{yb}=U$. Therefore, Charlie's observables $C_z=C_{0|z}-C_{1|z}$ satisfy $C_0=U\sigma_x U^\dagger$ and $C_1=U\sigma_z U^\dagger$.

\section{Bounding the sharpness parameter from noisy correlations}\label{AppB}
In order to bound the sharpness of Bob's instrument, consider first the witness $W_\text{AB}$. Using the notations from the previous Appendix, we have that 
\begin{equation}
W_\text{AB}=\frac{1}{8}\left(4+|\vec{c}_0||\vec{m}_0|\hat{m}_0\cdot\hat{c}_0+|\vec{c}_1||\vec{m}_1|\hat{m}_1\cdot\hat{c}_1\right).
\end{equation}
We focus on the simplified case in which the sharpness parameter is the same in Bob's two settings, i.e.~$\eta\equiv |\vec{c}_0|=|\vec{c}_1|$. Re-arranging gives
\begin{equation}
\eta=\frac{8W_\text{AB}-4}{|\vec{m}_0|\hat{m}_0\cdot\hat{c}_0+|\vec{m}_1|\hat{m}_1\cdot\hat{c}_1}.
\end{equation}
To find the smallest possible $\eta$, we maximise the denominator. That corresponds to setting $\hat{m}_0\cdot\hat{c}_0=\hat{m}_1\cdot\hat{c}_1=1$ and $|\vec{m}_0|=|\vec{m}_1|=\sqrt{2}$. That gives the lower bound
\begin{equation}
\eta \geq \sqrt{2}\left(2W_\text{AB}-1\right).
\end{equation}

Consider now the witness $W_\text{AC}$. In the previous Appendix, we have shown that its optimal value for a given choice of $\eta\equiv |\vec{c}_0|=|\vec{c}_1|$ is upper bounded as follows 
\begin{equation}
W_\text{AC}\leq \frac{1}{2}+\frac{1}{4\sqrt{2}}\left(1+\sqrt{1-\eta^2}\right).
\end{equation}
Solving this inequality for $\eta$ gives
\begin{equation}
\eta \leq 2\sqrt{\left(2+\sqrt{2}-4W_\text{AC}\right)\left(2W_\text{AC}-1\right)}.
\end{equation}

\end{document}